\documentclass [12pt]{article}
\usepackage {graphicx}
\usepackage {longtable}

\begin{document}
\begin{center}
\textbf{Localization of gravitational energy and its potential to evaluation 
of hydrogen atom properties} 
\end{center}

\bigskip

\begin{center}
Jozef \v{S}ima and Miroslav S\'{u}ken\'{\i}k
\end{center}

\begin{center}
Slovak Technical University, Radlinsk\'{e}ho 9, 812 37 Bratislava, Slovakia
\end{center}

\begin{center}
e-mail: sima@chtf.stuba.sk
\end{center}

\bigskip

Abstract. Vaidya metric as an integral part of the Expansive
Nondecelerative Universe (ENU) model enables to localize the energy of
gravitational field and, subsequently, to find a deep
interrelationship between quantum mechanics and the general theory of
relativity. In the present paper, stemming from the ENU model,
ionisation energy and energy of hyperfine splitting of the hydrogen
atom, energy of the elementary quantum of action, as well as the
proton and electron mass are independently expressed through the mass
of the planckton, Z and W bosons and fundamental constants.

\bigskip

\begin{center}
INTRODUCTION 
\end{center}

\bigskip

The time when science will be able to offer a definite solution of
relationship between quantum mechanics and the general theory of
relativity (GTR) is still far-away. In our previous works [1, 2] it
has been documented that results and consequences stemming from the
GTR can be applied, apart from the macroworld phenomena of cosmology
and astrophysics, to microworld phenomena too. In addition to
differences between the principles governing quantum mechanics and
gravity, there are also common features of these fields of physics. In
unification of these fields, two main obstacles can be identified,
namely their different nature (the uncertainty principle as a leading
feature of quantum mechanics contrary precise results emerging from
the GTR) and inability of up-to-date used physical theories to
localize gravitational energy.

As for the localization of gravitational energy, three main streams of
opinions can be observed [3]: 1) gravitational field energy is
localizable but a corresponding "magic" formula for its density is to
be found; 2) it is nonlocalizable in principle; 3) it does not exist
at all since the gravity is a pure geometric phenomenon. It seems that
solution of this enigma lies in the metrics applied. It has be
evidenced by the ENU model that a metric involving changes in matter
due to its permanent creation must be used.  Utilization of Vaidya
metric [4] allowed to offer answers to several open questions, explain
some known facts in a new independent light, correct some opinions or
demostrate their limitations and unveiled deep mutual
interrelationships between natural phenomena [1, 2]. This paper is
aimed to provide further evidence of capability of the ENU model and
the GTR in solving the issues of microworld by offering a new approach
to a deeper understanding the ionisation energy of the hydrogen atom,
the energy of hyperfine splitting, and the energy of elementary
quantum of action.  Moreover, it is evidenced that the proton and
electron mass being themselves fundamental constants, can be expressed
through the mass of the planckton, Z and W bosons and other
fundamental constants.

\bigskip

\begin{center}
THEORETICAL BACKGROUND 
\end{center}

\bigskip

The Planck energy $E_{Pc} $

\begin{equation}
\label{eq1}
E_{Pc} = m_{Pc} .c^2 = \sqrt {\frac{{\hbar .c^{5}}}{{G}}} \cong 
10^{19}GeV
\end{equation}

\noindent
where $m_{Pc} $ ($2.176711 \times 10^{ - 8}kg$) is the planckton mass and 
$G ( 6.67259 \times 10^{ - 11}$ $kg^{ - 1}m^{3}s^{ - 2} )$ is 
the gravitational constant, is of fundamental importance for space structure 
and existence of the Universe [5]. This energy plays an important role in 
unifying the fundamental physical interactions. Planck quantities - energy, 
length ($1.616051 \times 10^{ - 35}$m) and time ($5.390563 \times 10^{ - 
44}$ s) - represent limits describing the initial phase of the Universe 
expansion.

In our previous work [1] the density of gravitational energy has been 
expressed within the first approximation using Tolman equation as

\begin{equation}
\label{eq2}
\varepsilon _{g} = - \frac{{R.c^{4}}}{{8\pi .G}} \cong - 
\frac{{3m.c^{2}}}{{4\pi .a.r^{2}}}
\end{equation}

\bigskip

\noindent
where $\varepsilon _{g} $ is the density of gravitational energy emitted by 
a body with the mass \textit{m} at the distance \textit{r}, \textit{R} 
denotes the scalar curvature (contrary to a more frequently used 
Schwarzschild metric, in the Vaidya metric \textit{R $ \ne $} 0 also outside 
the body) and $a$ is the gauge factor. In ENU model, the gravitational 
energy is both localizable and quantifiable which enables to bridge quantum 
mechanics and the GTR and to provide answers to some problems which have 
been unsolvable up-to-now. Some of the potentials of ENU model are 
demonstrated further.

We are postulating that the gravitational energy of planckton in Compton 
volume $V_{C} $(determined by integration of the gravitational field density 
of planckton over Compton volume for any ``elementary'' particle) is equal 
to the rest energy $E_{0} $ of that particle. In other words, the particle 
energy is created by the planckton gravitational energy. This hypothesis can 
be expressed by relation

\begin{equation}
\label{eq3}
E_{0} = \int {\left| {\varepsilon _{g\left( {Pc} \right)}}  \right|dV_{C} 
\cong m_{Pc} .c^{2}\frac{{\lambda _{C}} }{{a_{\left( {T} \right)}} }} 
\end{equation}

\noindent
in which $a_{\left( {T} \right)} $ is the gauge factor and $\lambda _{C} $ 
is the Compton wavelength. In the above relation (\ref{eq3}), the Compton wavelength 
$\lambda _{C} $ relates to the mass $m$ of a given particle [6]

\begin{equation}
\label{eq4}
\lambda _{C} = \frac{{\hbar} }{{m.c}}
\end{equation}

\noindent
and the gauge factor $a_{\left( {T} \right)} $ relates to the specific time 
when it held

\begin{equation}
\label{eq5}
k.T \cong m.c^{2}
\end{equation}

\noindent
where $T$ is the temperature of the Universe. In the period starting at the 
beginning of the Universe expansion and finishing at the end of radiation 
era it had to hold [7]

\begin{equation}
\label{eq6}
E \approx T \approx a^{ - 1/2}
\end{equation}

Taking (\ref{eq5}) and (\ref{eq6}) into account, relation (\ref{eq7}) is obtained

\begin{equation}
\label{eq7}
a_{\left( {T} \right)} = \frac{{m_{Pc}^{2} .l_{Pc}} }{{m^{2}}}
\end{equation}

\noindent
where Planck length $l_{Pc} $ is defined as

\begin{equation}
\label{eq8}
l_{Pc} = \left( {\frac{{G.\hbar} }{{c^{3}}}} \right)^{1/2}
\end{equation}

A substitution of (\ref{eq4}) - (\ref{eq8}) into (\ref{eq3}) leads to
identity that is important for understanding a mutual relation between
the gravitational and inertial masses and, moreover, proves internal
consistency of the used procedure.

\bigskip

\begin{center}
RESULTS AND DISCUSSION
\end{center}

\bigskip

This paper is aimed at verification of a more general validity of (\ref{eq3}). In 
case of an efford to find the properties of a particle, Compton volume must 
be replaced by another volume characteristic for this particle. To verify 
the justification of a broader usability of (\ref{eq3}), the hydrogen atom was 
chosen. Its ``volume'' given by the Bohr radius (bearing in mind 
consequences of the uncertainty principle) and composition are known. In 
addition, its energy parameters (ionization energy, fine structure constant 
and hyperfine splitting) have been experimentally measured, their values 
(belonging to the most precisely determined values in the whole physics) are 
known and this allows to confront the results obtained within our approach 
with the reality.

To solve the problems associated with verification of (\ref{eq3}), it was necessary 
to determine the gauge factor corresponding to influence of both 
gravitational and electromagnetic forces. Further, solutions for some 
different cases are offered.

\bigskip

\textit{1. Ionization energy of the hydrogen atom}

\bigskip

In unification of electromagnetic and weak interactions, Z and W bosons play 
the crucial role. Gravitational influence of Z and W bosons on their 
surroundings initiated manifests itself just when their Compton wavelength 
became equal to their effective gravitational radius [1, 2]. At that time, 
gauge factor (denoted here as $a_{I\left( {H} \right)} $) reached the value

\begin{equation}
\label{eq9}
a_{I\left( {H} \right)} = \frac{{\hbar ^{2}}}{{G.m_{ZW}^{3}} }
\end{equation}

\noindent
where $m_{ZW} $ is the mean mass of Z and W bosons (the actual masses being 
$1.434 \times 10^{ - 25}$kg and $1.6262 \times 10^{ - 25}$ kg, 
respectively). Introducing (\ref{eq9}) to (\ref{eq3}) and substituting Compton wavelength in 
(\ref{eq3}) for the Bohr radius $r_{H} $ (52.917706 pm) [8] we obtain

\begin{equation}
\label{eq10}
I_{H} \cong \frac{{m_{Pc} .c^{2}.r_{H} .G.m_{ZW}^{3}} }{{\hbar ^{2}}}
\end{equation}

The left side of (\ref{eq10}) represents the ionisation energy $I_{H} $ of the 
hydrogen atom. Calculation using numerical values of the right-side members 
leads to the value of 14.0 eV that is very closed to the experimental value 
13,6 eV.

\bigskip

\textit{2. The mass of the electron}

\bigskip

The mass of the electron belongs to fundamental constants without explaining 
its value. In this part it is shown of how this inertial mass depends on 
other parameters and fundamental constants. The electron mass is a parameter 
present in expression of both the ionization energy (\ref{eq11}) and Bohr radius 
(\ref{eq12}) of the hydrogen atom [8, 9]

\begin{equation}
\label{eq11}
I_{H} = \frac{{m_{e} .e^{4}}}{{32\pi ^{2}.\varepsilon _{o}^{2} .\hbar ^{2}}} 
= \frac{{\alpha _{e}^{2} .m_{e} .c^{2}}}{{2}}
\end{equation}

\begin{equation}
\label{eq12}
r_{H} = \frac{{4\pi .\varepsilon _{o} .\hbar ^{2}}}{{m_{e} .e^{2}}} = 
\frac{{\hbar} }{{\alpha _{e} .m_{e} .c}}
\end{equation}

In (\ref{eq11}) and (\ref{eq12}), $\alpha _{e} $ (7.29735 x 10$^{-3}$) is the dimensionless 
fine structure constant (related to the spin-orbit coupling of the 
electron), $\varepsilon _{o} $ (8.854187816 x 10$^{-12}$ kg$^{-1} $m$^{-3 
}$s$^{4} $A$^{2}$) is the vacuum permittivity, $m_{e} $ (9.109534 x 
10$^{-31}$ kg) and $e$ (1.60217733 x 10$^{-19} $A.s) are the electron mass 
and charge, respectively. (To be exact, the reduced electron mass should be 
taken into account, the difference of the masses is, however, customarily 
omitted) Stemming from (\ref{eq10}), (\ref{eq11}) and (\ref{eq12}), relation (\ref{eq13}) correctly 
expressing the electron mass is obtained

\begin{equation}
\label{eq13}
m_{e} = \sqrt {\frac{{2m_{ZW}^{3}} }{{\alpha _{e}^{3} .m_{Pc}} }} 
\end{equation}

Calculation leads to the electron mass of $9.2029 \times 10^{ - 31}$ kg that 
is very closed to the actual value. Expressing $m_{ZW} $as [2]

\begin{equation}
\label{eq14}
m_{ZW} \cong \sqrt {\frac{{\hbar ^{3}}}{{g_{F} .c}}} 
\end{equation}

($g_{F} $ is the Fermi constant reaching $1.41 \times 10^{ - 62}J.m^{3}$), 
and its subsequent substitution into (\ref{eq13}) allows to express the electron 
mass through fundamental physical constants only as

\begin{equation}
\label{eq15}
m_{e} \cong \sqrt[{4}]{{\frac{{4\hbar ^{9}}}{{\alpha _{e}^{6} .m_{Pc}^{2} 
.g_{F}^{3} .c^{3}}}}}
\end{equation}

\bigskip

Equation (\ref{eq15}) reveals a deep interrelationship between the electron mass 
(being itself a fundamental constant) and other fundamental constants, and 
unveils the reason of its value.

\bigskip

\textit{3. Energy of hyperfine splitting in the hydrogen atom}

\bigskip

A further gauge factor (denoted as $a_{hf} $) is related to the time of 
initial gravitational influence of proton on its environment. Similarly to 
(\ref{eq9}) it holds ($m_{p} $ being the proton mass)

\begin{equation}
\label{eq16}
a_{hf} = \frac{{\hbar ^{2}}}{{G.m_{p}^{3}} }
\end{equation}

Substituting (\ref{eq16}) to (\ref{eq3}) we obtain

\begin{equation}
\label{eq17}
E_{hf} \cong \frac{{m_{Pc} .c^{2}.r_{H} .G.m_{p}^{3}} }{{\hbar ^{2}}}
\end{equation}

\bigskip

It must be connected to the energy of hyperfine splitting since this 
quantity depends on magnetic moments and thus, in turn, on the inertial 
masses of proton and electron. The electron mass is included in $r_{H} $ and 
calculation based on (\ref{eq17}) leads to

\begin{equation}
\label{eq18}
E_{hf} \cong 2.9 \times 10^{ - 24}J
\end{equation}

The above value is approximately 3 times higher than the experimentally 
determined energy of hyperfine splitting (a consequence of the interaction 
of magnetic momentum of the proton and electron in the hydrogen atom [10] 
being 1420.4057518 MHz or in wavelength units approximately 21 cm). It 
should be pointed out, however, that the calculation was performed only on 
the level of first approximation. Moreover, the value in (\ref{eq18}) is still much 
more precise than that obtained by means of usually used classical formula 
(\ref{eq19})

\begin{equation}
\label{eq19}
E_{hf} \cong \alpha _{e}^{2} .I_{H} .\frac{{m_{e}} }{{m_{p}} }
\end{equation}

\textit{4. The mass of the proton}

\bigskip

Using (\ref{eq17}) and (\ref{eq19}), a simplified relation determining
the mass of the proton emerges

\begin{equation}
\label{eq20}
m_{p} \cong \sqrt[{4}]{{\frac{{\alpha _{e}^{5} .m_{Pc} .m_{e}^{3}} }{{2}}}}
\end{equation}

In analogy with the electron mass, substituting $m_{e}$ in
(\ref{eq20}) for (\ref{eq15}) an expression for the interrelationship
of the proton mass and fundamental constants emerges exhibiting the
reason for its value as well as for the ratio of the electron mass and
proton mass. It is worth mentioning that to obtain precise (as closed
to experimental as possible) values, instead of simplified expressions
their exact (and more complex) forms must be used and, in addition,
the level of accord in calculated and measured values depends also on
the accuracy of values of all members in the relevant relations.

\bigskip

\textit{5. Energy of elementary quantum of action}

\bigskip

Due to the fact that the energy of gravitational field exerted by electron 
to its surroundings is lower than the critical gravitational density, the 
electron does not exert any gravitational effect on its surroundings at the 
time being [2]. Gravitational influence of electron will be observable when 
the gauge factor (denoted here as $a_{g\left( {e} \right)} $) reaches the 
value

\begin{equation}
\label{eq21}
a_{g\left( {e} \right)} = \frac{{\hbar ^{2}}}{{G.m_{e}^{3}} }
\end{equation}

Substituting (\ref{eq21}) into (\ref{eq3}) relation describing the energy of elementary 
quantum of action

\begin{equation}
\label{eq22}
E_{eq} \cong \frac{{m_{Pc} .c^{2}.r_{H} .G.m_{e}^{3}} }{{\hbar ^{2}}} \cong 
4.8 \times 10^{ - 34}J
\end{equation}

\noindent
is obtained. The above value corresponds (and is closed at a unit frequency) 
to the Planck constant $h$.

\bigskip

\begin{center}
Conclusions
\end{center}

\bigskip

Utilization of the background of the ENU model, as a model enabling to 
localize the energy of gravitational field, helped to unveil mutual 
relationships between some fundamental physical constants associated with 
the hydrogen atom, and to provide correct values of both its energy 
parameters and inertial mass of its constituents. The results presented in 
this contribution clearly documented the applicability of the ENU model in 
acting as a bridge connecting the macroworld phenomena described by the GTR 
and the quantum mechanical realm of particles. Verification of the postulate 
stating that the energy of any particle is created by the planckton 
gravitational energy using the hydrogen atom as an example, and excellent 
agreement of calculated and experimental values of its parameters should be 
taken as a starting point for further investigation and seeking of 
interrelationships and common features of the macroworld and microworld.

\bigskip

\begin{center}
References
\end{center}

\bigskip

1. J. \v{S}ima, M. S\'{u}ken\'{\i}k, \textit{General Relativity and Quantum Cosmology, 
}Preprint gr-qc/9903090

2. M. S\'{u}ken\'{\i}k, J. \v{S}ima, J. Vanko, \textit{General Relativity and Quantum 
Cosmology,}-Preprint gr-qc/0010061

3. V. Ullmann, Gravity, Black Holes and the Physics of Time-Space, 
Czechoslovak Astronomic Society, CSAV, Ostrava, 1986 (in Czech)

4. P.C. Vaidya, Proc. Indian Acad. Sci., A33 (1951) 264

5. V.L. Ginsburg, Sovremennaya astrofizika, Nauka, Moscow, 1970

6. G. Wataghin in Cosmology, Fusion and Other Matters (F. Reines, ed.), Adam 
Hilger, Ltd., London, 1972, p. 54

7. K. Krane, Modern Physics, 2nd ed., Wiley, New York, 1996, p. 535

8. F.J. Blatt, Modern Physics, McGraw/Hill, New York, 1992

9. F.L. Pilar, Elementary Quantum Chemistry, 2nd ed., McGraw-Hill, New York, 
1990

10. B. G. Baym, Lectures on Quantum Mechanics, W.A. Benjamin, Amsterdam, 
1969

\end{document}